\documentclass[times, 10pt,twocolumn]{article}
\usepackage{latex8}
\usepackage{times}
\usepackage[english]{babel}
\usepackage[latin1]{inputenc}
\usepackage[T1]{fontenc}
\usepackage{graphicx}
\usepackage{latexsym}
\usepackage{amsmath}
\usepackage{amssymb}
\def\gradi{\ensuremath{^\circ}}
\def\newblock{}

\def\limbo{\hbox{L${}^2$imbo}}

{\bf}{\rm}
{\bf}{\rm}
{\bf}{\rm}
{\bf}{\rm}
\def\qed{\relax\ifmmode\hskip2em \Box\else\unskip\nobreak\hskip1em $\Box$\fi}
\def\coldot{,} 
 {\catcode`\.=\active\gdef.{$\egroup\setbox2=\hbox to \dimen0 \bgroup$\coldot}}
\def\rightdots#1{%
  \setbox0=\hbox{$1$}\dimen0=#1\wd0
  \setbox0=\hbox{$\coldot$}\advance\dimen0\wd0
  \setbox2=\hbox to \dimen0 {}%
  \setbox0=\hbox\bgroup\mathcode`\.="8000 $}
\def\endrightdots{$\hfil\egroup\box0\box2}

\newlength{\widthplusfourty}
\newcommand{\ReL}{\hbox{$\mathcal R\!\raise2pt\hbox{$\varepsilon$}\!\hbox{$\mathcal L$}$}}
\newcommand{\ReLinda}{\hbox{$\mathcal R\!\raise2pt\hbox{$\varepsilon$}\!\hbox{$\mathcal L$inda}$}}

\graphicspath{{png/}}
\newcommand{\SubSubSection}[1]{\subsubsection{\hskip -1em.~#1}}
\begin{document}
\title{A System Structure for Adaptive Mobile Applications}
\date{}
%
\author{Vincenzo De Florio and Chris Blondia\\
  University of Antwerp\\
  Department of Mathematics and Computer Science\\
  Performance Analysis of Telecommunication Systems group\\
  Middelheimlaan 1, 2020 Antwerp, Belgium, \emph{and}\\
  Interdisciplinary institute for BroadBand Technology\\
  Gaston Crommenlaan 8, 9050 Ghent-Ledeberg, Belgium}

\maketitle
\begin{abstract}
A system structure for adaptive mobile applications is introduced
and discussed, together with a compliant architecture
and a prototypic implementation. A methodology is also
introduced, which exploits our structure to
decompose the behavior
of non stable systems into a set of quasi-stable scenarios. Within each of these
scenarios we can exploit the knowledge of the available QoS figures
to express simpler and better adaptation strategies.
\end{abstract}
\Section{Introduction}
The class of distributed applications (DA) has nowadays become too large to comprise
all its members without ambiguity.
Today, when discussing DA, it has become important to specify explicitly
an attribute, namely stability: is it possible or not to assume that
both DA and their environments are stable? In other words---is it possible to rely
on the fact that, with a reasonably high probability, the system under consideration
and the environments where it will be moved to will not change their characteristics?
If this is the case we call the corresponding DA as stable.

Why is stability so important? If a DA is stable, it is possible to rely on
some useful properties:

\begin{enumerate}
\item The system model is simpler and easier to capture, and the fault model
is well defined and less complex to deal with.
\item Events such as system partitioning do not occur very often.
As a consequence, designing mechanisms to recover transparently from those events
is worthwhile and provides the users and the developer with a virtual tightly coupled system.
In particular, if connections restore themselves tolerating partitionings,
then connection-orientation is an effective and valuable communication paradigm~\cite{Dav98}.
\item Furthermore, it is possible to provide 
transparency of distribution and lower level technicalities by means of
middleware support.
\end{enumerate}

Unfortunately the assumption of stability can not always be safely drawn:
in particular, the class of mobile distributed applications (MA) embodies
applications that by their own intrinsic nature are not stable. MA 
are applications for which it is more complex and difficult to enforce and guarantee
an agreed upon quality of service (QoS). Moreover, mobility implies
reduced size and, consequently, reduced computing power and available energy.
Those figures must not be made transparent---on the contrary, they need to be
monitored constantly and made available throughout the system architecture.
Any sub-system must be able to
access those figures to adjust its service to the current system and environmental conditions.

This said, we can draw now two observations:

\begin{enumerate}
\item Common solutions such as connection-orientation and transparency
are not adequate for MA~\cite{Dav98,Dav94,Her03,Katz94}.
Trying to enforce the connection-oriented communication model
despite the inherently many system partitionings experienced by non-stable
environments is a Sisyphean labor---both costly
and ineffective.
\item MA require two specific abilities---that of detecting the
various ``changes'' characterizing their system and surrounding environment,
and that of adapting accordingly their course in order to maximize
the ratio QoS over costs. The latter property is the so-called adaptability.
It is worth noting how transparency and adaptability are mutually incompatible.
\end{enumerate}

We can conclude that MA \emph{must\/} be structured after the above
abilities, and that in particular they require \emph{system support for adaptability}.

We present herein an architecture and a system structure for adaptive MA
addressing the above requirements systematically. Our paper is
structured as follows:
in Sect.~\ref{s:adapt} we deepen our discussion on adaptability
and its requirements.
Section~\ref{s:contrib} details our contributions in abstract terms.
Section~\ref{s:devel} describes a prototype implementation
and some results.
Section~\ref{s:voting} provides an example of how our approach may be used to improve
the perceived QoS of adaptive mobile services.
Section~\ref{s:end} concludes our discussion recalling
main contributions and current state of development.

\Section{Adaptability and its Requirements}\label{s:adapt}
As remarked, e.g., in~\cite{Dav98}, 
a truly extended use of mobile computing technologies
asks for effective software engineering techniques to design,
develop and maintain \emph{adaptive\/} applications, i.e.
applications that are prepared to continue the distribution of
their service despite the inherently
significant and rapid changes in their supporting infrastructure
and, in particular, in the quality-of-service (QoS) available
from their underlying communications channels. In other words,
applications meant for mobile environments must adapt in response
to internal changes; to changes in the location of the client software; and to changes
in the characteristics of the environment~\cite{Dav94,Katz94}.
Therefore, general mechanisms to facilitate adaptation are becoming an
important requirement for distributed systems platforms and
mobile software engineering.
%

Previous research has delineated the key services required by
such mechanisms: adaptation asks for (at least) QoS change detection
and QoS feedback and control~\cite{Dav95,Gar04,Katz96,Nob95}. Here we discuss these
requirements.

\SubSection{Detection of Changes}\label{s:detection}
A fundamental design goal of (stable) distributed system
is transparency. This means providing the illusion of a common
homogeneous communication and computation environment, which
facilitates distributed software development, maintenance, and
re-use. Unfortunately, distribution transparency means
hiding the environmental changes, which substantially
prohibits adaptation. On the contrary, adaptation calls for
explicit QoS information acquisition throughout the
end-system software \cite{Dav95,Katz96,Nob95}.
Mobile systems platforms must collate
and manage QoS information originating at
the communication layers and the end-systems.
Examples of such QoS information
include power availability, physical location, device
proximity and communications capabilities and costs~\cite{Dav98}.
Such information must be made available to the application and
control layers in order to drive the adaptation processes.

\SubSection{Feedback and Control}
Once a meaningful QoS change has occurred and has been detected, the next
step is to react to this event. An example of reaction could be, e.g.,
adapting the error protection scheme to the current state (this is supported,
for instance, by the TETRA system, which provides configurable variable
bit rate channels with multiple levels of forward error
protection).
Achieving, by proper feedback and control,
an optimal degree of information redundancy is another example.
Another one could be exploiting power information on host peripherals
triggering hardware power saving functionality as
appropriate.

\Section{An Architecture for Adaptive Mobile Applications}\label{s:contrib}
We propose herein an architecture for orchestrating run-time adaptation
of MA and a methodology to make profitable use of that architecture. In the following we
describe these two contributions.

\SubSection{Architecture}
Our architecture is structured after the requirements of adaptability sketched in Sect.\ref{s:adapt},
with a set of layers dealing with changes detection and publication and another layer managing
feedback and control. Changes detection and publication is inspired by the works
of Davies et al.~\cite{Dav98} and is reported in Sect.~\ref{s:contrib:detection}.
Our adaptation layer for feedback and control is described in following Sect.~\ref{s:contrib:fandc}.

\SubSubSection{Change detection}\label{s:contrib:detection}
Change detection is realized by components that we call ``change detectors'' (CD). Similarly to
the QoS agents of \limbo~\cite{Dav98} and the probes of Rainbow~\cite{Gar04},
CD are communication or end-system
components that ``publish'' information such as power availability, energy availability,
physical location, device proximity and communications capabilities and costs.
Whatever the originator, the published information takes the form of
Linda-like tuples~\cite{CaGe89a}, called ``change notifications'' (CN).

Exploiting Linda in MA is not a new concept---the reader may refer for instance
to~\cite{Busi03,Dav98,Her03,Pic99,Tol03}.

A CN is emitted through a Linda operator such as \texttt{out} or \texttt{eval},
or through a special function called \texttt{evalp}. The latter produces
what we call a ``live tuple with passive snapshot'' (the corresponding tuple is
``alive''~\cite{CaGe89a} but, each time its value changes, a passive copy is updated).

As in plain Linda systems, the emitted or updated tuple reaches a repository
called Tuple Space (TS) and is maintained by a distributed component called Tuple Space Manager
(TSM). 
Tuples can be queried by any component in the system.
Tuples can be withdrawn or updated only by their original producers.

The key difference between other approaches and ours is that the reception of
a tuple by the TSM not only results in the publishing of the corresponding CN;
in addition, it
triggers feedback and control, as shown in next sub-section.

The TSM also keeps track of the current structure and state of the system, the user
application, and the TS.

\SubSubSection{Feedback and Control}\label{s:contrib:fandc}
Following each insertion, the TSM awakens a component that we call
Adaptation Strategies Interpreter (ASI). This is a virtual machine that interprets 
small programs written in a simple scripting language called Ariel.

\begin{figure}[t]
\includegraphics[width=0.50\textwidth]{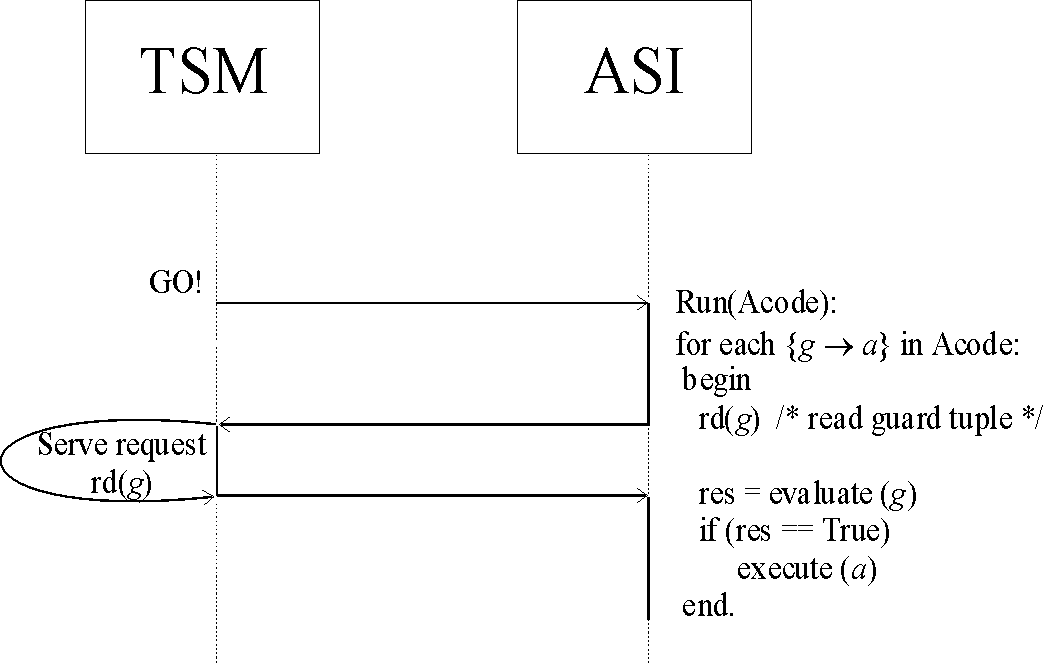}
\caption{Execution of an Ariel program.}
\label{f:ua-ts-ra}
\end{figure}

The only control structure offered by Ariel is Guarded Actions~\cite{Dijk76}.
Guarded actions are conditioned actions---actions that are executed only when their 
pre-condition (the guard) is evaluated as true. Conditions evaluation is carried
out by checking the contents of the TS (see Fig.~\ref{f:ua-ts-ra}).
Actions trigger system-wide control by issuing
other tuples or executing control commands. Commands can, e.g., disassociate one or more
station, change remote tasks priorities, modify protocol parameters, or can be associated
to user methods. Their scope range from local to global.

\begin{figure}[t]
\hskip-15pt%
\includegraphics[width=0.50\textwidth]{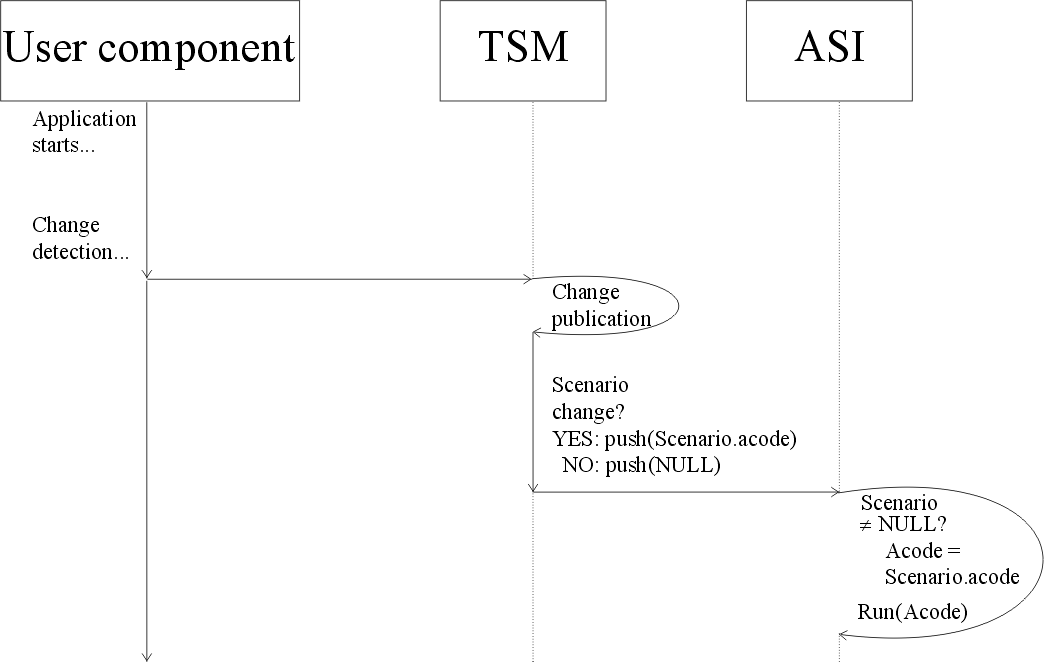}
\caption{Scenario detection and execution.}
\label{f:ua-db-ra}
\end{figure}

Each Ariel program takes care of dealing
with a particular \emph{scenario}, the latter being a set of predefined
QoS values. For instance, a scenario could be
``\emph{For all Stations: Energy $>$ 40\% AND CPU\_Usage $<$ 75\%}''.
The default scenario is called Otherwise, and is the one the MA is in when no other scenario can be
selected. Adaptation is enforced by detecting a scenario change and loading ASI with the corresponding
Ariel program. 

It is worth observing that doing like just described means \emph{decomposing the behavior
of a non stable system into a set of quasi stable scenarios}. Within each of these
scenarios we can exploit the known QoS figures to express simpler and better strategies.

For example, if we can rely on the fact that the system is currently not subjected to malicious
attacks, we can select a weaker authentication protocol thus assuaging the system of the
corresponding overhead. Such a scenario may be expressed, e.g., as follows: 
``\emph{For all Stations: Failed\_Attempts\_Per\_Sec $=$ 0}''.
Other scenarios may e.g. put on the foreground the observed reliability of the channels.
Corresponding strategies could enforce an optimal degree of information redundancy such that
the required QoS be guaranteed while both maximizing the channel throughput and
minimizing the costs of its usage.

\SubSection{Methodology}
The choice of which scenarios to consider can be done in different ways. The one we suggest
herein is based on so-called Pareto ophelimity~\cite{Pareto}, which is the generalization of
optimality for multi-dimensional design evaluation spaces. A point in one such space is
said a Pareto point if all other points are worse in at least one of the
dimensions\footnote{Pareto states that the optimum allocation of the resources of a
                    society is not attained so long as it is possible to make at least one
                    individual better off in his own estimation
                    while keeping others as well off as before in their own estimation.}.
Pareto optimality can be used as e.g. in~\cite{Wong03} 
for energy-aware design-time task scheduling of dynamic telecommunication and multimedia systems,
or in~\cite{L3} for performance- and energy-aware dynamic data types transformation and refinement.

The proposed methodology encompasses the following steps:
\begin{itemize}
\item Monitoring the system response and QoS under various configurations of, e.g.,
available energy, or bandwidth, or local or overall CPU usage.
\item Computing a Pareto curve trading off, e.g., power availability and observed throughput.
\item Defining the scenarios corresponding to the points in the Pareto curve. 
\item Writing an Ariel program for each of these points and associating it to its scenario.
\item Detecting at run-time ``where the system is'' with respect to the Pareto curve and,
      accordingly, pushing the corresponding Ariel program onto ASI (see Fig.~\ref{f:ua-ts-ra}).
\end{itemize}
Under the assumption of correct detection of QoS scenarios and that of correct Ariel designs, these
Ariel programs are selected and can realize a system-wide orchestration of user-defined adaptation
strategies.
%

\Section{Prototype}\label{s:devel}
This section describes a prototypic implementation of our adaptation architecture.
We distinguish a run-time part and a compile-time part---they are described respectively in
Sect.~\ref{s:arch:rt} and Sect.~\ref{s:arch:ct}.
Section~\ref{s:run} shows some preliminary results.

\begin{figure}[t]
\centerline{\includegraphics[width=0.55\textwidth]{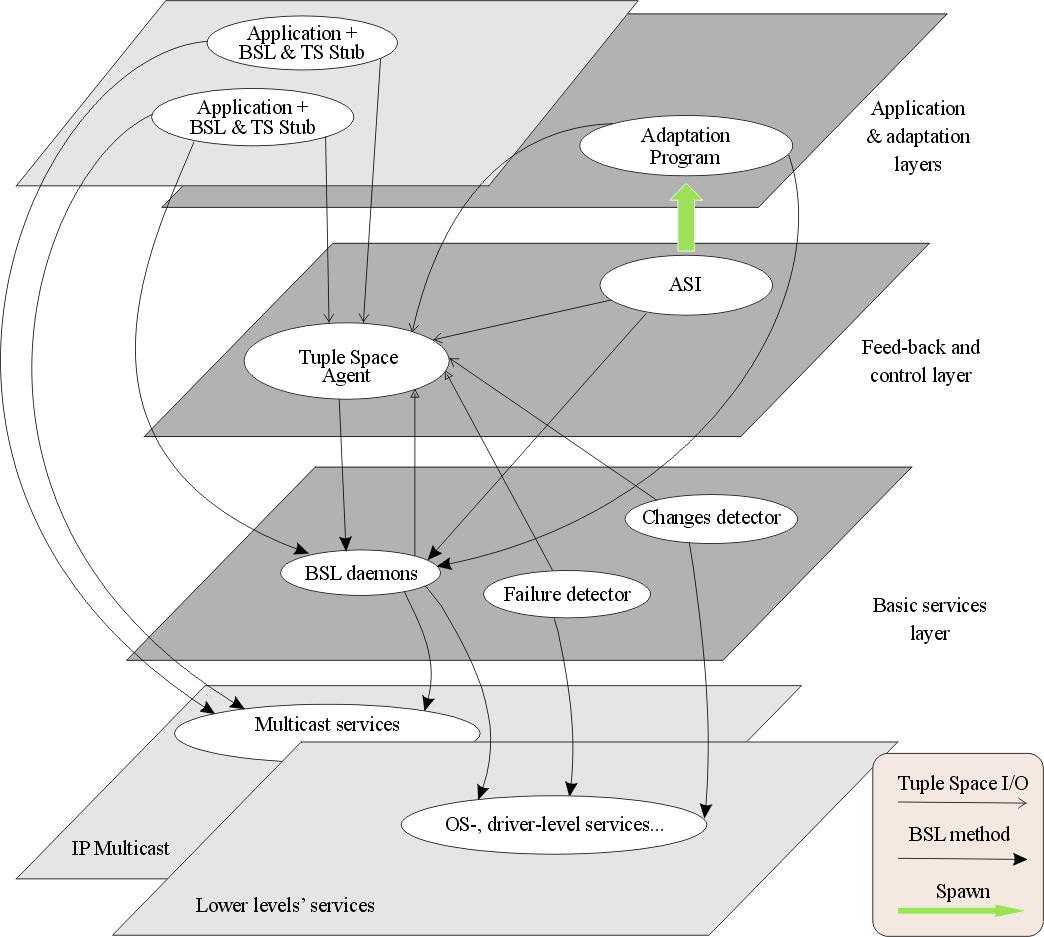}}
\caption{The picture portrays in grey the three layers of the node architecture.
White layers on bottom are the system communication and control layers. On top,
the application layer is also portrayed in white. Note that application components need to
be compiled with a BSL and a TS stub.}\label{f:node}
\end{figure}

\SubSection{Run-time Components}\label{s:arch:rt}
Figure~\ref{f:node} provides a view to its structure on a single station. We call this
the node architecture (NA). The same structure is to appear on each participating station.

NA consists of three layers:
\begin{description}
\item[The Basic Services Layer] (BSL) exports common services from the 
underlying communication and control layers. These services include
asynchronous, connectionless group communication and remote task
creation/management/termination. Similarly to
\limbo, two daemon processes are used for this tasks. This layer also hosts a failure
detector~\cite{ChTo43} and one or more change detectors.
\item[The Feedback-and-Control Layer] hosts the local agents of the TSM and of the ASI.
The state of these two components is cyclically checked by the BSL failure detector.
\item[The Adaptation Layer] runs the Ariel program associated with the currently
detected scenario.
\end{description}

The user applications can be written on any programming language. No restrictions
apply to the adopted communication model---in particular applications are not restricted
to use Linda operations, though special control and monitoring is achieved when the
application tasks do make use of either Linda or BSL methods for communication. In particular,
an Ariel script can only isolate those tasks that make use of either Linda or BSL methods.

In the current prototypic implementation, scenarios are ``hard-coded'' into the TSM.

\SubSection{Compile-time Components}\label{s:arch:ct}
\SubSubSection{Ariel}
Ariel was originally developed as a programming language
for distributed applications with dependability
requirements~\cite{DF00}.

Ariel is a language to code corrective actions when certain state changes occur and are
detected. The main characteristics of Ariel is that it is expressed in a separate
application layer: it can be thought of as a separate thread of execution that
asynchronously gets activated at each relevant state change. When this occurs
the Ariel program gets executed. Execution is carried out by interpreting a
pseudo-code called a-code. The a-code interpreter is the ASI component
introduced in Sect.~\ref{s:contrib:fandc}. 

As already mentioned, Ariel programs are a list of if-then-else constructs, possibly nested,
which represent guarded actions (GA). GA are executed in the order of appearance in the
Ariel source code.
Actions deal with coarse-grained entities
of the application (stations, access points, tasks, and groups of tasks)
while guards query the current state of those entities, as
it is stored in the TS. Figure~\ref{f:ua-ts-ra} provides the
general scheme of execution of an Ariel program.

For a the general syntax of Ariel and a list of its guards and actions
we refer the reader to~\cite{DF00}.

An important
consequence of the adoption of this strategy is that
the functional code and the adaptation
code are \emph{distinct}: the former implements
the user tasks while the latter is given by a proper
coding of the adaptability actions.
This allows to decompose the design process into two
distinct phases. When the interface between the two
``aspects'' is simple and well-defined, this provides
a way to control the design complexity and has positive
relapses on development and maintenance times and costs.

Summarizing, Ariel
allows to express the application software as
two separate codes: the functional code and the
a-code. The former deals with the specification
of the functional service whereas the latter is the
description of the measures that need to be
taken in order to perform some corrective actions,
such as ordering the modification of some
key parameter like, for instance, the
code redundancy used in data transmission, or which
software process needs to be appointed to
a given sub-task.
The specification of these corrective actions
is done by the user in an environment other
than the one for the specification of
the functional aspects.

This separation still
holds at run-time, since the executable code
and the a-code are physically distinct.
This strict separation between the two
aspects may allow to ``trade'' at run-time the actual
set of adaptation actions to be executed, as described
in Sect.~\ref{s:contrib}.

\SubSubSection{Other Tools}
A number of ancillary tools have been designed around Ariel. In particular
an Ariel translator, called ``art,'' changes Ariel scripts into a-code sets
coded as integer triplets in statically allocated arrays.

Another tool, called ``rcodenv,'' has been developed to bundle together in
one large header file all the a-code sets corresponding to the various
scenarios. The resulting file needs to be compiled with ASI.

\begin{figure}[t]
\hskip-5pt%
\includegraphics[width=0.52\textwidth]{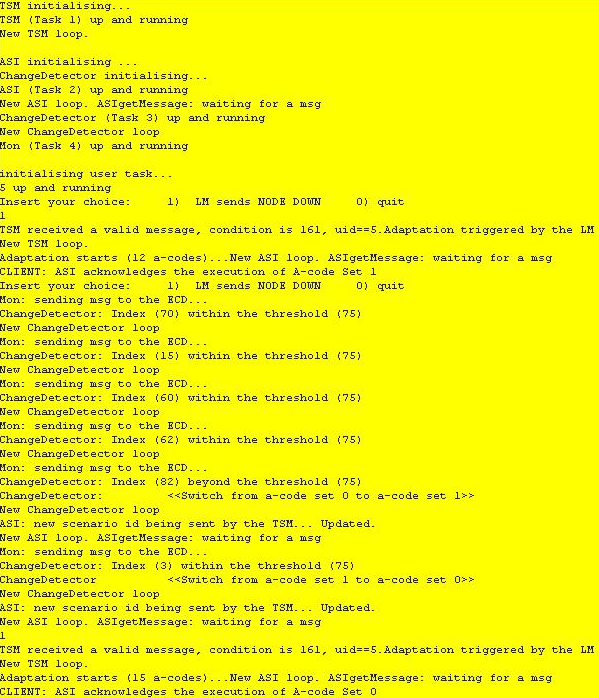}
\caption{A system with two scenarios.}
\label{f:run1}
\end{figure}

\begin{figure}[t]
\hskip-5pt%
\includegraphics[width=0.52\textwidth]{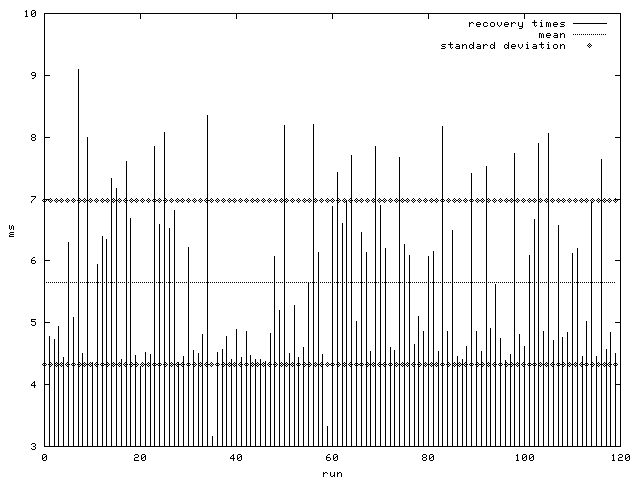}
\caption{Elapsed times for the execution of an adaptation program by the ASI
virtual machine.}
\label{f:times}
\end{figure}

\SubSection{Preliminary results}\label{s:run}
Here we report on some preliminary results obtained with a
prototype system running on a cluster of Linux workstations.
Figure~\ref{f:run1} shows a system with two scenarios:
``CPU\_Usage $<$ 75\%'' and Otherwise. A CD reports the current level
of the CPU while a dummy application called LM is used to trigger adaptation
reporting that a node is down.
Two adaptation programs are selectable---though in this case they both
do the same action: broadcasting a system-wide alarm.
Figure~\ref{f:times} reports the execution times for program 0, consisting of
15 pseudo-codes.

\Section{Adaptive Voting Sensors}\label{s:voting}
We propose herein an example of how our approach may be used to improve
the perceived QoS of adaptive mobile services.

In what follows our target service is remote monitoring of patients 
through Body Area Networks (BANs) of wireless sensors.
Permanent monitoring and logging of vital signs is achieved by means of a set of mobile, compact units that
continuously transfer and publish the value of a pre-defined set of vital parameters between a patient's location and the clinic or the doctor in charge. Quality of this service here is defined as 
the service's trustworthiness \emph{and\/} cost-effectiveness: 
\begin{description}
\item[R1:] (Hard) guarantees are required so that, whenever the patient is in need, the system
is to trigger a system alarm (e.g., dispatching medical care to the patient).
\item[R2:] (Soft) guarantees are required, such that no false alarm is triggered when the patient is not in real
need---the latter to reduce the service costs. 
\end{description}

The above contradicting requirements are not easily reconcilable---\textbf{R1} would require the system alarm to be
triggered whenever any one of the sensors alerts or disconnects, while \textbf{R2} would commit to system alarm only
after several of those events. Trade-offs are possible, of course, and can be easily expressed, e.g., through
$m$-out-of-$n$ majority voting: if at least $m$ sensors out of the $n$ currently reachable sensors are
alerting then we trigger system alarm.

Whatever the choice of $m$, this approach may turn up to be too unflexible and hence be perceived
as unsatisfactory by the users. We propose the following alternative method based on our approach:
\begin{enumerate}
\item For each class of patient, a base of parameters is isolated, including e.g., heartbeats per minute,
body temperature, or arterial pressure.
\item From that base we derive a set of sensors to be adopted in the remote monitoring system.
\item We isolate a set of scenarios with respect to the above parameters (one such scenario
for instance may be ``heartbeat = 70, temperature = 38$\gradi$C, arterial pressure = 120'').
\item We partition the scenarios with respect to seriousness of symptoms and we define an Ariel program
for each class.
\end{enumerate}
Each Ariel program $p$ may, e.g., compute an $m(p)$-out-of-$n$ voting. Doing so, a different trade-off
between \textbf{R1} and \textbf{R2} would be chosen depending on the derived symptoms.

\Section{Conclusions and Future Work}\label{s:end}
We introduced a
system structure for adaptive mobile applications based on 
decomposing the behavior
of a non stable system into a set of quasi stable scenarios. Within each of these
scenarios we can exploit the known QoS figures to express simpler and better strategies.
Foreseen future work includes augmenting our prototype,
exercising it on various case studies, and analyzing its performance.


\bibliographystyle{latex8}

\end{document}